\documentclass[aps,prl,twocolumn,superscriptaddress]{revtex4}

\usepackage[dvips]{graphicx}
\usepackage{color}
\usepackage{bm}
\usepackage{amssymb}

\usepackage[utf8]{inputenc}

\begin{document}

\title{Frustrated spin-1/2 molecular magnetism in the mixed-valence antiferromagnets Ba$_3M$Ru$_2$O$_9$ ($M$~$=$~In, Y, Lu)}

\author{D.~Ziat}
\affiliation{Institut Quantique and Département de Physique, Université de Sherbrooke, 2500 boul. de l'Université, Sherbrooke (Québec) J1K 2R1 Canada}

\author{A.~A.~Aczel}
\email{aczelaa@ornl.gov}
\affiliation{Quantum Condensed Matter Division, Oak Ridge National Laboratory, Oak Ridge, TN 37831, USA}

\author{R.~Sinclair}
\author{Q.~Chen}
\author{H.~D.~Zhou}
\affiliation{Department of Physics and Astronomy, University of Tennessee, Knoxville, Tennessee, 37996-1200, USA}

\author{T.~J.~Williams}
\affiliation{Quantum Condensed Matter Division, Oak Ridge National Laboratory, Oak Ridge, TN 37831, USA}

\author{M.~B.~Stone}
\affiliation{Quantum Condensed Matter Division, Oak Ridge National Laboratory, Oak Ridge, TN 37831, USA}

\author{A.~Verrier}
\affiliation{Institut Quantique and Département de Physique, Université de Sherbrooke, 2500 boul. de l'Université, Sherbrooke (Québec) J1K 2R1 Canada}

\author{J.~A.~Quilliam}
\email{jeffrey.quilliam@usherbrooke.ca}
\affiliation{Institut Quantique and Département de Physique, Université de Sherbrooke, 2500 boul. de l'Université, Sherbrooke (Québec) J1K 2R1 Canada}

\date{\today}

\begin{abstract}

We have performed magnetic susceptibility, heat capacity, muon spin relaxation, and neutron scattering measurements on three members of the family Ba$_3M$Ru$_2$O$_9$, where $M$~$=$~In, Y and Lu. These systems consist of mixed-valence Ru dimers on a triangular lattice with antiferromagnetic interdimer exchange. Although previous work has argued that charge order within the dimers or intradimer double exchange plays an important role in determining the magnetic properties, our results suggest that the dimers are better described as molecular units due to significant orbital hybridization, resulting in one spin-1/2 moment distributed equally over the two Ru sites. These molecular building blocks form a frustrated, quasi-two-dimensional triangular lattice. Our zero and longitudinal field $\mu$SR results indicate that the molecular moments develop a collective, static magnetic ground state, with oscillations of the zero field muon spin polarization indicative of long-range magnetic order in the Lu sample. The static magnetism is much more disordered in the Y and In samples, but they do not appear to be conventional spin glasses.

\end{abstract}


\maketitle

\section{I. Introduction}

The 6H-perovskites, with formula Ba$_3MA_2$O$_9$, have provided fertile ground for recent discoveries in frustrated quantum magnetism. Materials in this family with magnetic $M$-sites have been shown to exhibit quantum spin liquid behavior, in particular 6HB-Ba$_3$NiSb$_2$O$_9$~\cite{Cheng2011,Quilliam2016,Fak2016} and Ba$_3$IrTi$_2$O$_9$~\cite{Dey2012BITO} while others, Ba$_3$CuSb$_2$O$_9$~\cite{Zhou2011,Nakatsuji2012,Quilliam2012bcso} and Ba$_3$ZnIr$_2$O$_9$~\cite{Nag2016}, exhibit possible quantum spin-orbital liquids. Furthermore, Ba$_3$CoSb$_2$O$_9$ has allowed for some of the first studies on the magnetization process of a truly triangular spin-1/2 antiferromagnet~\cite{Shirata2012,Zhou2012Co,Susuki2013,Koutroulakis2015,Quirion2015}. The flexibility of this crystal structure means that we are also at liberty to include magnetic 4$d$/5$d$ transition metal $A$-site ions and thereby study spin dimers distributed on a triangular lattice with significant spin-orbit coupling and orbital hybridization. In the case of the ruthenates Ba$_3M$Ru$_2$O$_9$, where $M^{3+}$ is non-magnetic, one obtains a triangular lattice of magnetic, mixed-valence Ru dimers. A total of seven electrons occupy each dimer and this leads to the possibility of charge, orbital and spin degrees of freedom. 

For analogous 3$d$ transition metal-based dimer systems with more than two electrons per dimer, Hund's coupling is usually dominant and therefore needs to be treated before turning to intersite effects such as electron hopping and the interdimer Coulomb interaction. However, recent theoretical \cite{Streltsov2014, Streltsov2016} and experimental \cite{Kimber2012} work has shown that this approach can break down in some 4$d$ and 5$d$ transition metal-based dimer systems, where Hund's coupling is expected to be significantly weaker due to the spatially-extended $d$-orbitals. This more complicated regime may be realized in the Ba$_3M$Ru$_2$O$_9$ family, as any simple picture based on dominant Hund's coupling cannot describe all of the known magnetic properties of the Ru dimers.

\begin{figure*}
\begin{center}
\begin{tabular}{ll}
\hspace{-0.1in} \includegraphics[width=3in]{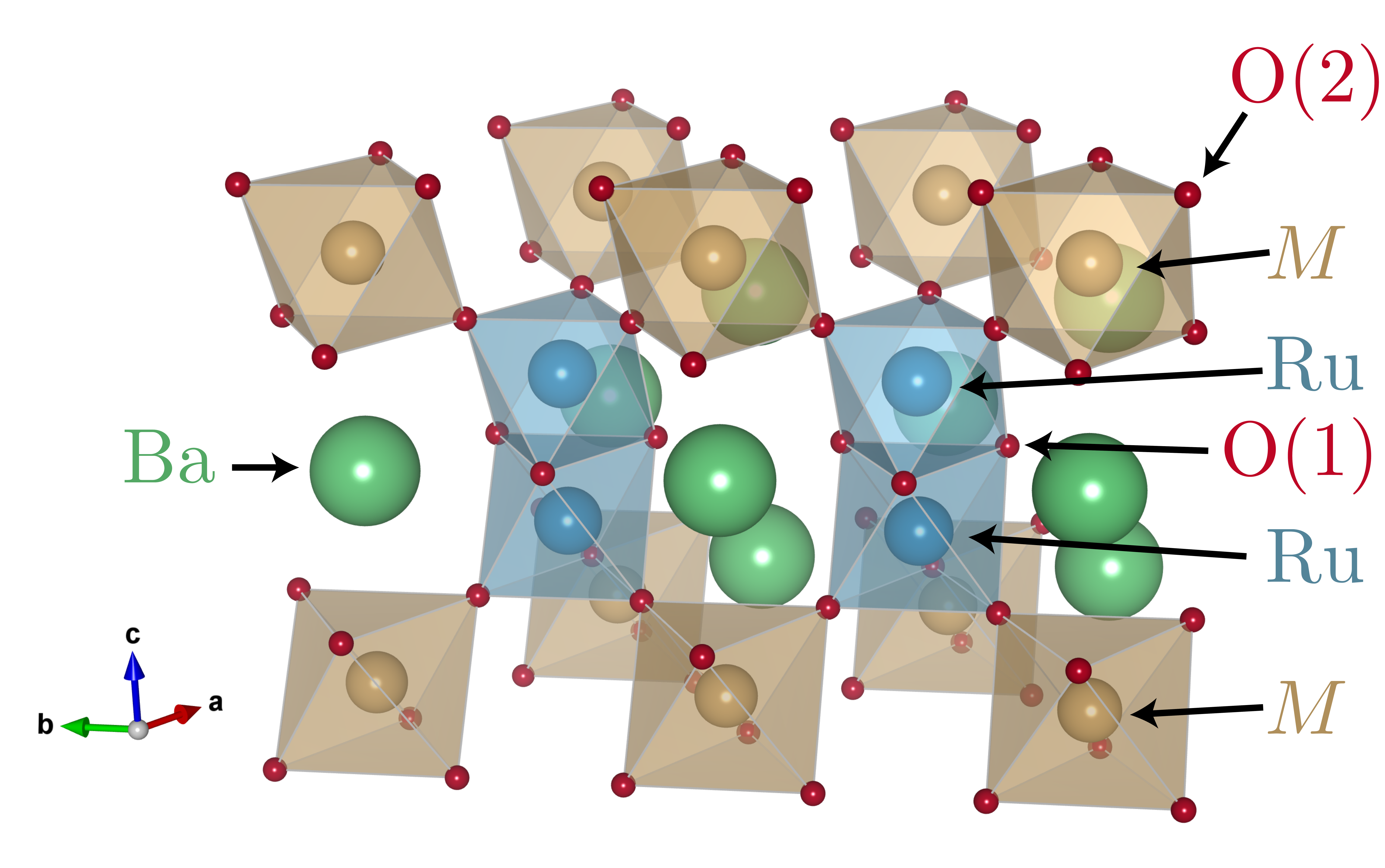}  \hspace{0.2in} &
\hspace{0.25in}\includegraphics[width=2.5in]{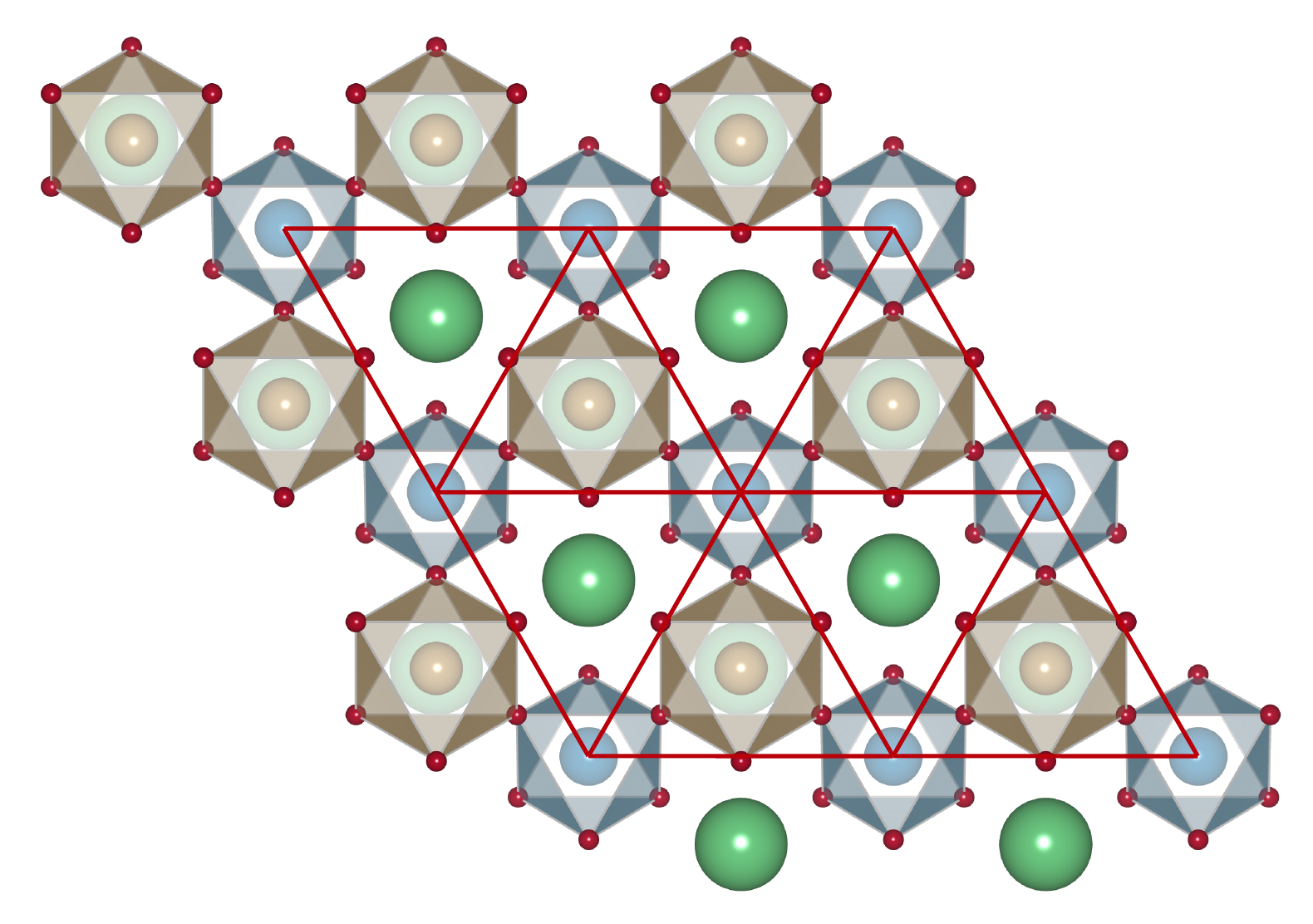} \\
\vspace{-1.8in}\\
 (a) & (b)  \\
 \vspace{1.4in} \\
\end{tabular}
\caption{\label{Structure} A portion of the crystal structure of Ba$_3M$Ru$_2$O$_9$, specifically using parameters for the $M=$Y sample, showing one plane of Ru-Ru dimers. (a) A view perpendicular to the $c$-axis showing the stacking of Ru ions to form dimers. (b) A view parallel to the $c$-axis, showing the triangular arrangement of Ru-dimers.}
\end{center}
\end{figure*}

More specifically, two different magnetic ground states for the Ru dimers in Ba$_3M$Ru$_2$O$_9$ have been proposed previously that are consistent with dominant Hund's coupling. Doi \emph{et al.}~\cite{Doi2002} first assumed that all seven electrons were localized at particular Ru sites, which leads to Ru$^{4+/5+}$ charge order within the dimers and antiferromagnetic intradimer exchange. They argued that the latter should produce dimers with a magnetic ground state of total spin $S$~$=$~1/2, which could explain the loss of effective magnetic moment with decreasing temperature in their $M$~$=$~In, Y and Lu magnetic susceptibility measurements and an entropy release of $R \ln(2)$ in their specific heat data at the low temperature magnetic phase transitions ($T_m$~$=$~4.5, 4.5, and 9.5~K for In, Y, and Lu). 

However, their model fails to explain the very different, monotonic susceptibility in the $M =$ La sample, as the intradimer exchange interaction would have to change dramatically, from strongly antiferromagnetic to strongly ferromagnetic with only a tiny modification of the crystal structure. Even if that were possible, the model would imply $S=5/2$ dimers in the La compound which would lead to much larger values of susceptibility than are measured. Furthermore, subsequent neutron diffraction measurements of the Y system found no evidence for the required charge ordering within the dimers down to 2 K \cite{Senn2013}. For these reasons, the magnetic ground state of the Ru dimers has also been discussed more recently in the context of molecular double exchange \cite{Senn2013}, but this simple model cannot explain the non-monotonic $T$-dependence of the magnetic susceptibility, the low-$T$ entropy release in the specific heat data, and the small ordered moment sizes for the Y and La systems found in neutron diffraction. 

This means that there is currently no comprehensive understanding of the magnetic ground states for single Ru dimers in the Ba$_3M$Ru$_2$O$_9$ family. The collective magnetic ground states of these materials may also be interesting in their own right, as the interdimer interactions are likely frustrated due to the triangular lattice geometry of the dimers. For these reasons, we have used magnetic susceptibility, heat capacity, muon spin relaxation ($\mu$SR), and neutron scattering to investigate both the single dimer and collective magnetic properties of the $M$~$=$~In, Y and Lu systems.

\section{II. Experimental details}

The polycrystalline samples of Ba$_3M$Ru$_2$O$_9$ ($M$~$=$~In, Y and Lu) studied here were prepared by the standard solid state reaction method. Appropriate amounts of BaCO$_3$,  In$_2$O$_3$/Y$_2$O$_3$/Lu$_2$O$_3$ (Y$_2$O$_3$ and Lu$_2$O$_3$ were pre-dried at 980 $^\circ$C overnight), and RuO$_2$ were mixed in agate mortars, compressed into pellets, and annealed for 20 hours in air at temperatures of $900^\circ$C, $1200^\circ$C and $1300^\circ$C, respectively. Magnetic susceptibility and specific heat measurements were performed using \emph{Quantum Design MPMS} and \emph{PPMS} systems. The DC magnetic susceptibility was measured with a magnetic field of 1~kG. AC susceptibility measurements were also performed at various frequencies (from 333 Hz to 9999 Hz) to look for evidence of spin freezing.

Neutron powder diffraction (NPD) was performed with polycrystalline Ba$_3M$Ru$_2$O$_9$ ($M$~$=$~In, Y and Lu) using the HB-2A powder diffractometer of the High Flux Isotope Reactor (HFIR) at Oak Ridge National Laboratory (ORNL). The Lu sample was loaded in a vanadium can, and the data was collected at $T$~$=$~1.5~K with a neutron wavelength of 1.54~\AA~and a collimation of 12$'$-open-12$'$. The In and Y samples were loaded in aluminum cans, and the data was collected at $T$~$=$~3.5~K with a neutron wavelength of 1.54~\AA~and a collimation of 12$'$-21$'$-6$'$. 

Complementary elastic neutron scattering measurements were performed on the fixed-incident-energy triple-axis spectrometer HB-1A of HFIR at ORNL, using the same polycrystalline samples. A series of two pyrolytic graphite (PG) crystal monochromators provided the fixed incident energy $E_i$ of 14.6 meV and two-highly oriented PG filters were placed in the incident beam to remove higher order wavelength contamination. A PG analyzer crystal was located before the single He-3 detector for energy discrimination. A collimation of 40$'$-40$'$-40$'$-80$'$ resulted in an energy resolution at the elastic line of $\approx$ 1 meV. The elastic scattering was measured at 1.5~K for all three samples, with higher temperature background data collected at 20 K for the Lu system and 10 K for the In and Y systems.

Inelastic neutron scattering (INS) measurements were collected on the direct-geometry time-of-flight chopper spectrometer SEQUOIA of the Spallation Neutron Source (SNS) at ORNL, using the same polycrystalline samples loaded in aluminum cans. Spectra were collected at a variety of temperatures by operating in high flux mode (elastic resolution of $\sim$ 4 \% $E_i$) with $E_i$~$=$~50 and 100 meV. The monochromatic incident beam was obtained by using a Fermi chopper rotating at a frequency of either 180 or 240 Hz for $E_i$~$=$~50 and 100~meV respectively. The background from the prompt pulse was removed with a $T_0$ chopper operating at 90~Hz. An empty aluminum can was measured in identical experimental conditions for a similar counting time. The resulting spectra were subtracted from the corresponding sample spectra after normalization with a vanadium standard to account for variations of the detector response and the solid angle coverage. This procedure ensured that temperature-independent scattering was removed from the spectra before applying the appropriate Bose corrections to calculate $f(Q)^2 \chi''(Q, \omega)$, where $\chi''(Q, \omega)$ is the imaginary part of the dynamic magnetic susceptibility and $f(Q)$ is the magnetic form factor. 

Muon spin relaxation measurements were performed at TRIUMF, Canada on the M20 beam line with the LAMPF spectrometer and a He-flow cryostat. Samples were encapsulated in Ag-coated mylar adhesive and suspended between copper supports in the path of the muon beam where they were cooled by helium vapour to as low as $\sim 2$ K. This style of sample mount and a veto counter behind the sample allow us to almost completely eliminate any background asymmetry.  Measurements were taken in zero-field (ZF), longitudinal field (LF), and weak transverse field (TF) geometries using forward and backward positron counters to determine the asymmetry, $a(t) = (n_B - \alpha n_F)/(n_B + \alpha n_F)$. $\alpha$ is determined with weak transverse field measurements in the paramagnetic phase and $a(t)$ is divided by the initial asymmetry to obtain the muon polarization, $P(t)$. 

 
 \begin{table}[htb]
\begin{center}
\caption{Structural parameters for Ba$_3M$Ru$_2$O$_9$ ($M$~$=$~In, Y, and Lu) extracted from the refinements of the $\lambda$~$=$~1.54~\AA~neutron powder diffraction data. The lattice constants and bond distances are in \AA, and the bond angles are in degrees.} 

\begin{tabular}{l l l l l}
\hline 
\hline
$B'$ & In (3.5 K) & Y (3.5 K) & Lu (1.5 K) & La (11 K) \cite{Senn2013} \\
\hline
$a$ & 5.7947(1) & 5.8565(1) & 5.8436(1) & 5.9492 \\  
$c$ & 14.2738(2) & 14.4589(1) & 14.3978(2) & 14.9981 \\
Ba$_2$ $z$ & 0.9116(2) & 0.9075(1) & 0.9084(2) & 0.8909 \\ 
Ru $z$ & 0.1611(1) & 0.1632(1) & 0.1620(1) & 0.16556 \\ 
O$_1$ $x$ & 0.4874(5) & 0.4879(4) & 0.4887(5) & 0.4873 \\
O$_2$ $x$ & 0.1712(4) & 0.1758(2) & 0.1741(3) & 0.17889 \\
O$_2$ $z$ & 0.4150(1) & 0.4124(1) & 0.4138(1) & 0.40471 \\
R$_\mathrm{wp}$ & 8.82~\% & 6.27~\% & 6.18~\% & 6.66~\% \\ 
Ru-O$_1$ & 2.001(3) & 2.009(2) & 2.019(2) & 2.030 \\
Ru-O$_2$ & 1.956(2) & 1.936(1) & 1.947(2) & 1.909 \\ 
Ru-Ru & 2.538(3) & 2.511(2) & 2.533(3) & 2.533 \\ 
Ru-O$_1$-Ru & 78.8(1) & 77.4(1) & 77.7(1) & 77.2 \\ 
\hline\hline
\end{tabular}
\end{center}
\end{table}

\section{III. Search for static charge order}
It is important to understand the magnetic ground state of a single Ru dimer before moving on to a discussion of these materials' collective magnetic properties. As shown in Fig.~\ref{Structure}(a), each Ru site is in an octahedral oxygen environment, and the Ru dimers form via face-sharing octahedra. It is well-known that all three materials crystallize in the space group $P 6_3/m m c$ at room temperature, which ensures that both Ru sites forming a dimer are crystallographically-equivalent due to the crystal symmetry. However, static charge order is a distinct possibility for these materials upon cooling due to the mixed Ru$^{4+/5+}$ nominal valence, which has been found in isostructural systems with a mixed Ru$^{5+/6+}$ nominal valence such as Ba$_3$NaRu$_2$O$_9$ \cite{Kimber2012}. Neutron powder diffraction (NPD) is a sensitive probe to look for this effect, as one can investigate the $T$-dependence of the charge distribution in the dimers indirectly via Ru-O bond lengths. 

\begin{figure}
\centering
\includegraphics[width=3.35in]{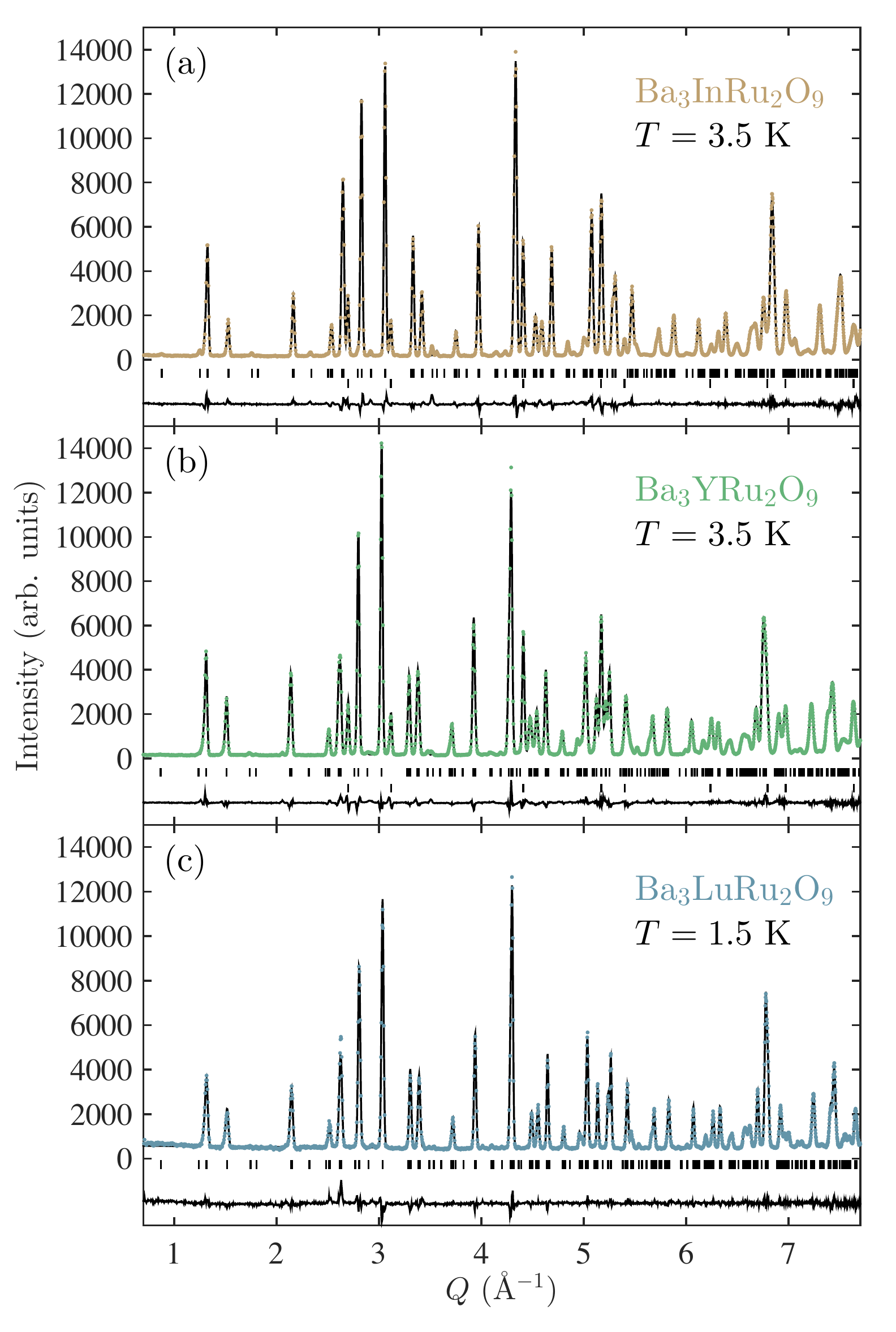}
\caption{\label{Diffraction} Neutron powder diffraction measurements with a wavelength of 1.54~\AA~for (a) Ba$_3$InRu$_2$O$_9$ (3.5 K), (b) Ba$_3$YRu$_2$O$_9$ (3.5 K), and (c) Ba$_3$LuRu$_2$O$_9$ (1.5 K). The corresponding structural refinements (black lines) are superimposed on the data points. The extra peaks in the In and Y patterns arise from the Al sample can.}
\end{figure}

Figure~\ref{Diffraction} shows NPD data collected using $\lambda$~$=$~1.54~\AA~for Ba$_3M$Ru$_2$O$_9$, with $T$~$=$~1.5~K for the Lu system and $T$~$=$~3.5~K for the In and Y analogs. Rietveld refinements were performed using FullProf \cite{Rodriguez1993}. In all cases, we find that the data is best refined in the room temperature $P 6_3/m m c$ space group with only one unique Ru site and no Ru/$M$ site mixing, and therefore we find no evidence for static charge ordering down to these temperatures. We also do not detect any magnetic Bragg peaks, which would be indicative of long-range magnetic order, in this data. Table I shows lattice constants, atomic fractional coordinates, and selected bond distances and angles extracted from the refinements. We note that our O$_2$~$z$~parameter for the Y system is significantly different from the value reported in Ref.~\cite{Senn2013}. Upon careful inspection their value appears to be somewhat unphysical~\cite{Footnote}

\section{IV. Molecular magnetism}

\begin{figure}
\begin{center}
\includegraphics[width=3.5in,keepaspectratio=true]{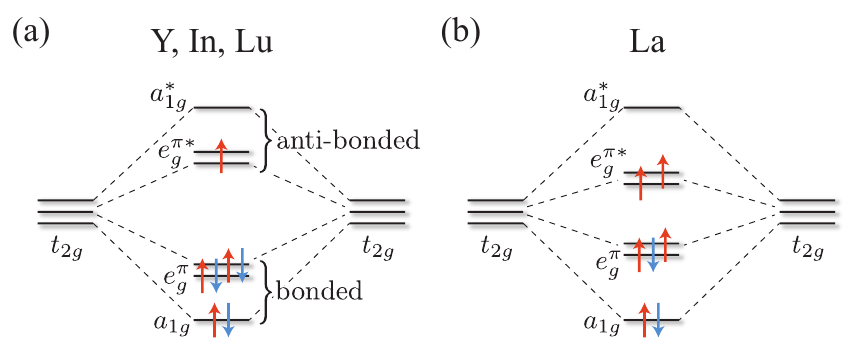}\\
\includegraphics[width=3.5in,keepaspectratio=true]{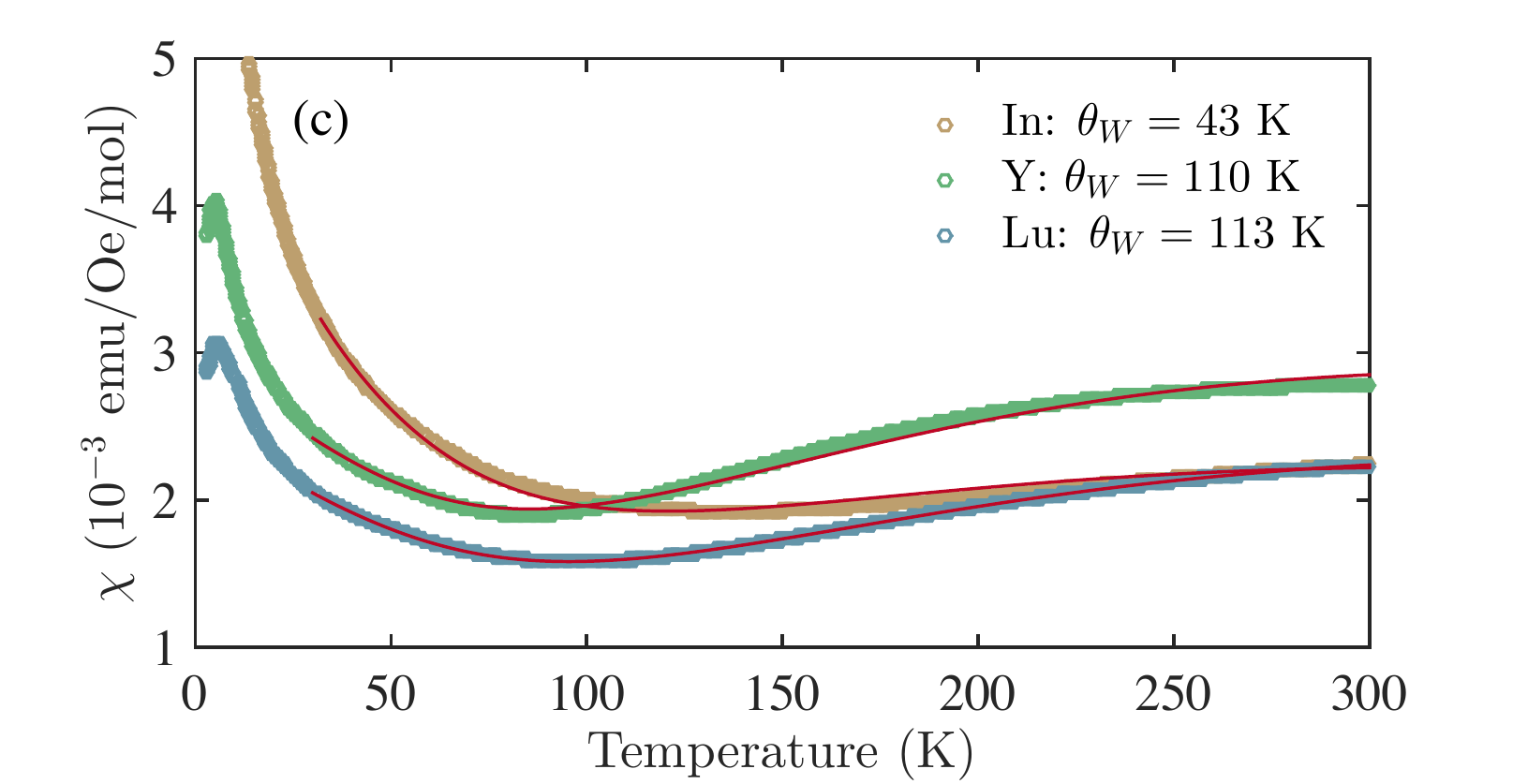}\\
\caption{(a) Energy-level and spin occupation diagram for a hybridized Ru$^{4.5+}$-Ru$^{4.5+}$ dimer with large bonding energy, which is likely to apply to the In, Y and Lu samples~\cite{Streltsov2016}. (b) Energy level diagram with lower bonding energy as expected to apply to the La sample. (c) Magnetic susceptibility of the In, Y and Lu samples, with the fits using Eq.~\ref{ChiModel}.\label{Orbitals} superimposed on the data.}
\end{center}
\end{figure}

Since there is no evidence for static charge order of the Ru dimers in Ba$_3M$Ru$_2$O$_9$ ($M$~$=$~In, Y and Lu), we now consider other possibilities for the single dimer ground states that are consistent with the known magnetic properties. We first revisit the DC magnetic susceptibility of these materials, as a satisfactory explanation for the complex $T$-dependence is still lacking. Our own results, shown in Fig.~\ref{Orbitals}(c), are very similar to previous work by Doi \emph{et al.}~\cite{Doi2002}. Between $\sim 100$~K and 300~K, $\chi$ is an increasing function of temperature ($d\chi/dT > 0$), suggestive of gapped spin excitations. Below $\sim 100$ K, however, $\chi$ becomes a decreasing function of temperature ($d\chi/dT<0$), \emph{i.e.} begins to resemble a Curie-Weiss law. A logical explanation for this non-monotonic behavior is a change in spin number with temperature. For example, the ground state of each Ru dimer may be a $S=1/2$ doublet with a relatively low-lying excited $S=3/2$ manifold (with energy $\Delta_1$). As $T>100$ K, we begin to populate the $S$~$=$~3/2 manifold, which naturally has a larger susceptibility. If we assume that there is also a $S$~$=$~5/2 manifold with higher energy $\Delta_2$, a minimal functional form for the susceptibility~\cite{Doi2002} can be written as
\begin{equation} 
\chi(T) = \frac{\mathcal{C}}{T + \Theta_W}\cdot \frac{1 + 10e^{-\Delta_1/k_BT} + 35e^{-\Delta_2/k_BT}}{1 + 2e^{-\Delta_1/Tk_B} + 3e^{-\Delta_2/k_BT}} 
\label{ChiModel}
\end{equation}
This equation accounts for interactions between dimers via the $\Theta_W$ term. Fits of this form provide an adequate description of the susceptibility data over a broad temperature range. Without fixing any parameters, these fits yield $\Delta_1 = 38.9(4)$ meV (In), 28.6(3) meV (Y) and 34.1(4) (Lu).  However, fits of this form are somewhat over parametrized and a more direct method for exploring the excitation spectrum is desirable.

\begin{figure}
\begin{center}
\includegraphics[width=3.35in,keepaspectratio=true]{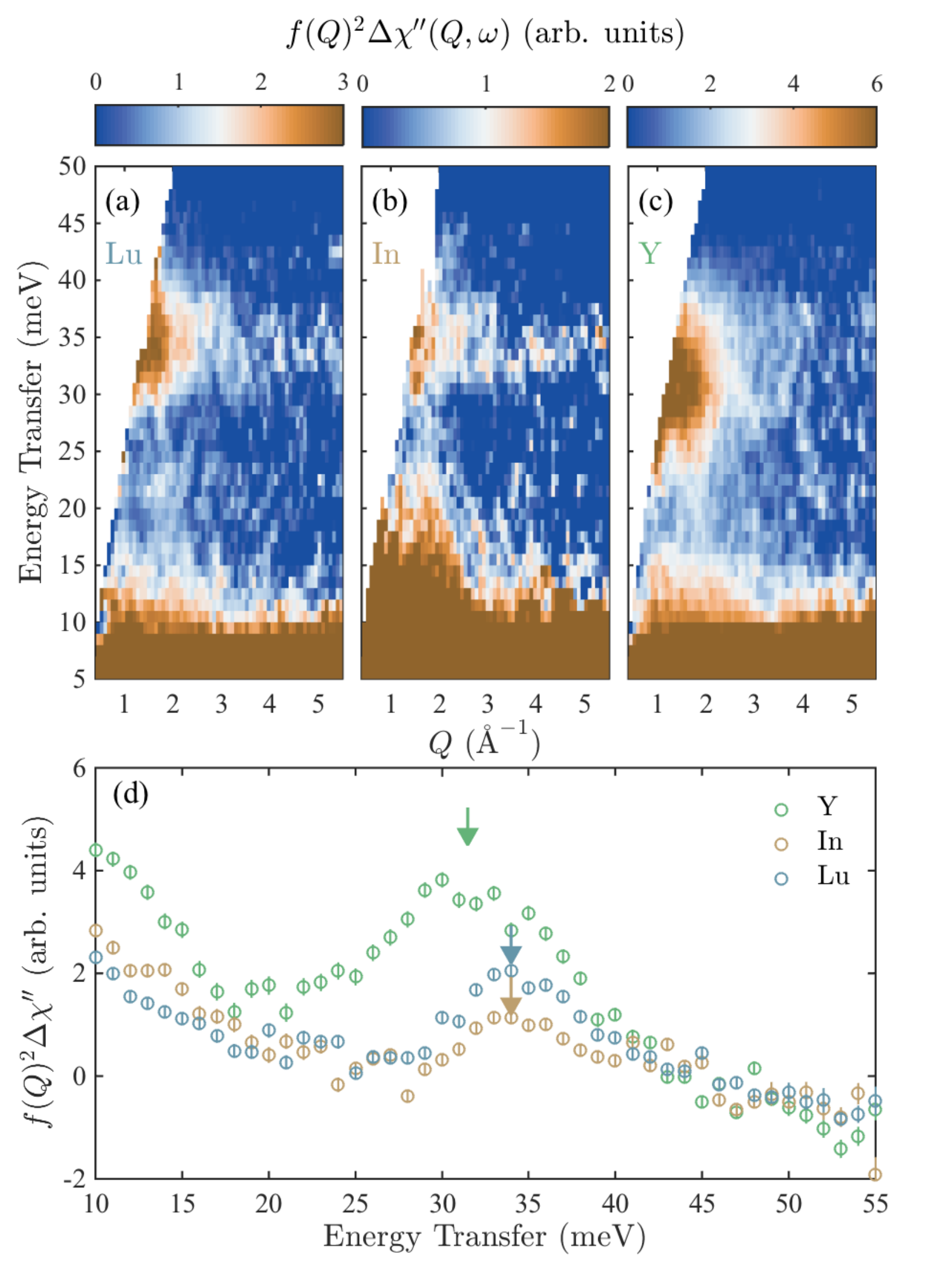}
\caption{$f(Q)^2 \Delta \chi'' = f(Q)^2 \chi''$(5 K) - $f(Q)^2 \chi''$(225 K) (arbitrary units) obtained with inelastic neutron scattering for (a) Ba$_3$LuRu$_2$O$_9$, (b) Ba$_3$InRu$_2$O$_9$ and (c) Ba$_3$YRu$_2$O$_9$ as a function of wave vector and energy transfer. (d) Cuts of $f(Q)^2\Delta \chi''$ integrated between $Q = 2$ and 2.5 \AA$^{-1}$. \label{Neutrons}}
\end{center}
\end{figure}

\begin{figure}
\begin{center}
\includegraphics[width=3.5in,keepaspectratio=true]{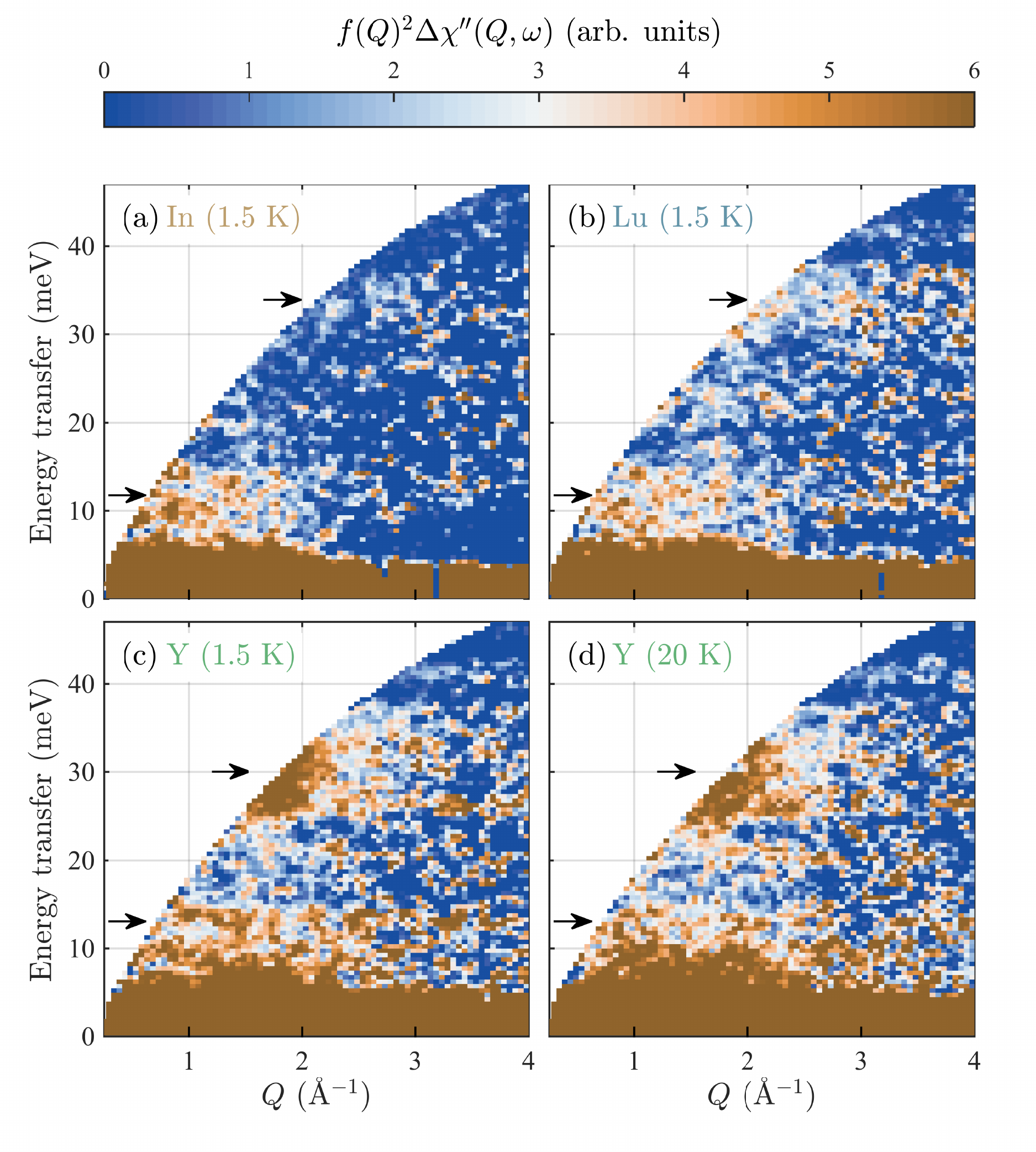}
\caption{$f(Q)^2 \Delta \chi'' = f(Q)^2 \chi''$($T$) - $f(Q)^2 \chi''$(225 K) (arbitrary units) obtained with inelastic neutron scattering as a function of wave vector and energy transfer, using a lower incident energy of $E_i = 50$ meV for (a) Ba$_3$LuRu$_2$O$_9$ at $T=1.5$ K (b) Ba$_3$InRu$_2$O$_9$ at $T=1.5$ K and Ba$_3$YRu$_2$O$_9$ at (c) $T=1.5$ K and (d) $T = 20$ K. \label{LowIncidentEnergy}}
\end{center}
\end{figure}

To this end, we have employed inelastic neutron scattering measurements, carried out on the SEQUOIA spectrometer with an incident energy of $E_i$~$=$~100~meV. We plot $f(Q)^2 \chi''(Q, \omega$) for Ba$_3M$Ru$_2$O$_9$ in Fig.~\ref{Neutrons}(a)-(c) as a temperature difference $f(Q)^2 \Delta \chi'' = f(Q)^2[\chi''(5\text{ K})-\chi''(225\text{ K})]$ to isolate the low-temperature magnetic scattering. Two dispersive magnetic modes are visible in the spectra of each system. The lower modes are located just above the elastic line and appear more clearly in the complementary $E_i$~$=$~50~meV datasets shown in Fig.~\ref{LowIncidentEnergy}. The finite dispersion of these modes likely arises from significant interdimer interactions. Constant-$Q$ cuts taken from the same datasets with an integration range of [2-2.5]~\AA$^{-1}$ are depicted in Fig.~\ref{Neutrons}(d). These cuts indicate that the  higher energy mode is centered about 34(1) meV, 31.5(1.5) meV, and 34(1) meV for the In, Y, and Lu systems respectively. These excitation energies correspond reasonably well to the values of $\Delta_1$ obtained from freely fitting the DC susceptibility. 

Ultimately, we have fitted the susceptibility data by fixing the values of $\Delta_1$ to those measured with our INS measurements, resulting in only 3-parameter fits and eliminating the over-parametrization problem. The Curie constants $\mathcal{C}$ obtained from this fitting give effective moment sizes, $\mu_\mathrm{eff}$, in the ground state manifold of $1.40(3)\mu_B$, $1.65(3)\mu_B$, and $1.53(3)\mu_B$ per dimer for the In, Y and Lu samples, respectively. These values are only slightly under the value of $1.73\mu_B$ expected for a free spin-1/2, and therefore this result is consistent with our proposal that a single dimer has a total spin $S$~$=$~1/2 ground state. The Weiss constants, $\Theta_W$, are found to be 43(3) K, 110(10) K and 113(2) K for the In, Y and Lu systems respectively, which are indicative of significant antiferromagnetic interdimer exchange. The $S=5/2$ state is found at $\Delta_2 = $ 81(1) meV (In), 72(1) meV (Y) and 80(1) meV (Lu). Despite the high energy of $\Delta_2$, these states cannot be ignored in the susceptibility fitting. 

This model includes a number of simplifications, most importantly that the Weiss constant, $\Theta_W$, is the same in all manifolds of total spin. This is counterintuitive since one would expect a higher total spin to yield a larger Weiss constant, all things being equal, since $\Theta_W = 2zJS(S+1)/3k_B$ where $z$ is the number of nearest neighbors. The success of this simplistic model, in which $\Theta_W$ is constant, therefore implies that the interaction strength, $J$, between dimers is smaller when they are excited into their $S$~$=$~3/2 or $S$~$=$~5/2 manifolds, compensating for the increase in spin number. 	

This single dimer picture supported by our susceptibility and INS measurements can be better understood by drawing on the work of Streltsov and Khomskii who have investigated the possibility of covalent bonds forming between 4$d$/5$d$ ions in various cases \cite{Streltsov2014, Streltsov2016}. For the current Ba$_3 M$Ru$_2$O$_9$ structure, one should consider the transition metal Ru ions in the strong crystal field regime. Since these ions are in an octahedral oxygen environment, this assumption leads to the usual low energy $t_{2g}$ orbitals and higher energy $e_g$ orbitals. A trigonal distortion, inherent to this family of materials crystallizing in the $P6_3/m m c$ space group, then splits the $t_{2g}$ orbitals into an $a_{1g}$ singlet and an $e_{g}^\pi$ doublet~\cite{Kugel2015}. The unique face-sharing octahedral geometry of two neighboring Ru sites is argued to produce strong orbital hybridization, with the $a_{1g}$ orbitals experiencing the largest bonding energy as shown in Fig.~\ref{Orbitals}(a) and (b). If the two Ru sites are close enough, then the $e_{g}^\pi$ orbitals can also participate in molecular bonding. The choice of magnetic ground state for a single dimer in a particular system depends critically on the ratio of the molecular bonding energy to Hund's coupling, as illustrated in Fig.~\ref{Orbitals}(a) and (b). In the present materials Ba$_3M$Ru$_2$O$_9$ with $M$~$=$~Y, In and Lu, the molecular bonding energy appears to be higher than Hund's coupling, and therefore the electrons prefer to occupy the $e_g^\pi$ bonding orbitals rather than the $e_g^{\pi\ast}$ anti-bonding orbitals. In other words, three covalent bonds form and one uncompensated electron is left over. This situation is illustrated in Fig.~\ref{Orbitals}(a).

This model suggests that the higher-energy dispersive modes observed in the INS spectra, shown in Fig.~\ref{Neutrons}(a)-(c) and highlighted in the cuts of Fig.~\ref{Neutrons}(d), can be assigned to electron transitions from bonding to antibonding molecular orbitals, which would cause the total spin of a dimer to change from $S$~$=$~1/2 to 3/2.  The origin of the lower energy INS modes can also be understood in the context of the molecular magnet model, as they may simply represent electron transitions between the antibonding orbitals shown in Fig.~\ref{Orbitals}(a). Any origin associated with collective magnetic ordering or spin freezing for these low energy modes can be ruled out as there was no significant change observed in their temperature-dependence between 1.5 and 20~K in complementary $E_i$~$=$~50~meV datasets, which is illustrated for Ba$_3$YRu$_2$O$_9$ in Fig.~\ref{LowIncidentEnergy}.

On the other hand, the magnetic susceptibility of Ba$_3$LaRu$_2$O$_9$ \cite{Doi2002} is consistent with a total spin $S$~$=$~3/2 dimer ground state, and a $S=1/2$ excited state, which implies that the molecular bonding energy is not as large and therefore only two covalent bonds form in this case, as illustrated in Fig.~\ref{Orbitals}(b). This also explains the much larger magnetic moment observed in neutron diffraction experiments~\cite{Senn2013}. It is natural to ask what structural parameter gives rise to this dramatic difference between the La sample and the In, Y, and Lu analogs studied here. Although there is no discernible correlation with Ru-Ru distance as shown in Table~I, the La sample does have a larger Ru-O(1) distance and smaller Ru-O(1)-Ru bond angle than the other materials. These parameters may play an important role in determining the molecular bonding energy of the $e_g^\pi$ orbitals, especially since the O$_1$ ions form the common octahedral face of the Ru$_2$O$_9$ units. As can be seen from Fig.~\ref{Orbitals}(b), a smaller bonding energy leads to the $S=3/2$ configuration expected for the La sample.

\section{V. Collective magnetic ground states}
\begin{figure}
\begin{center}
\includegraphics[width=3.3in,keepaspectratio=true]{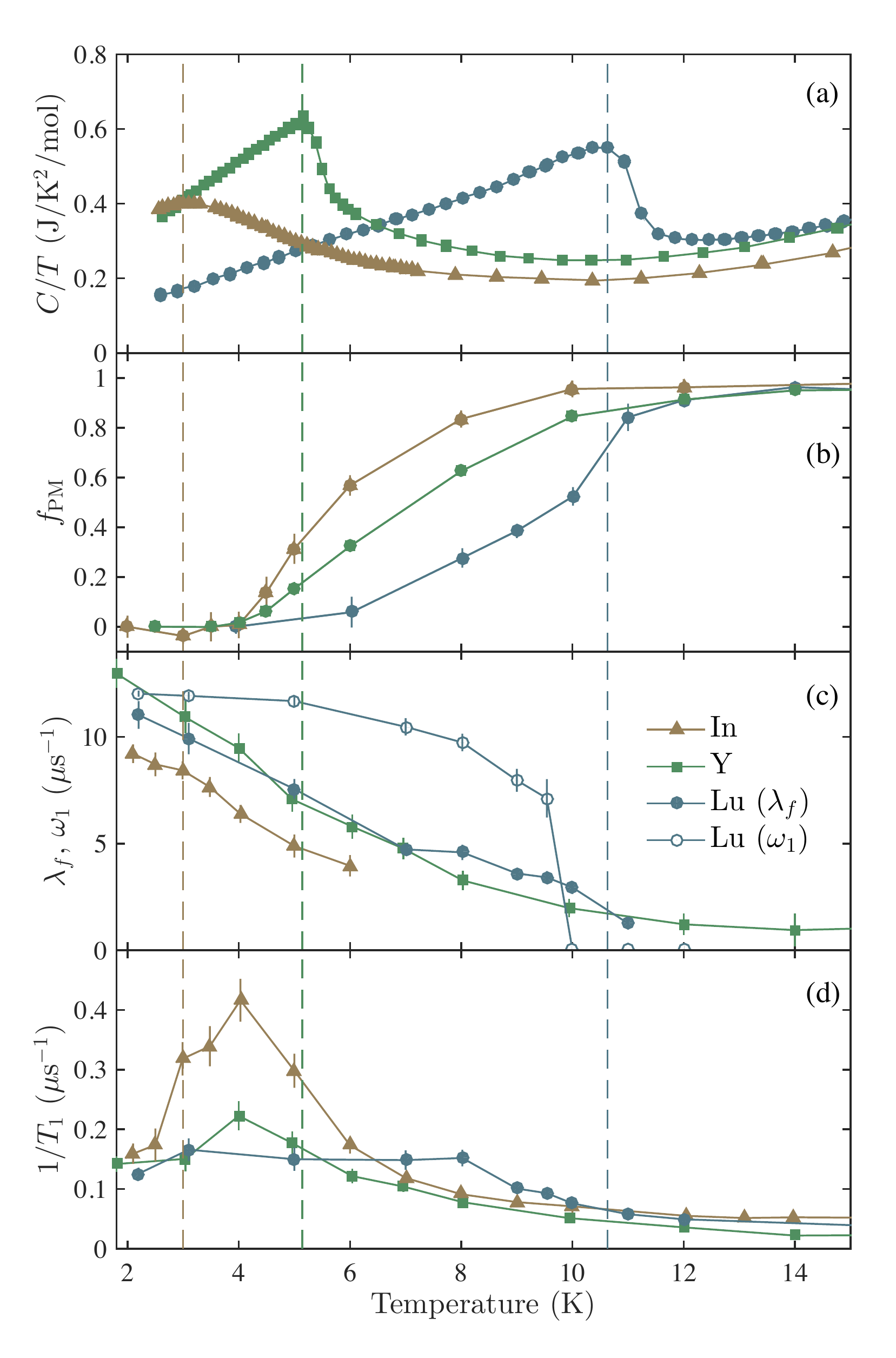}
\caption{(a) Specific heat ($C$) of the samples measured here with dashed lines identifying the low-$T$ anomalies as $T_m$. (b) The paramagnetic fraction of the samples as a function of temperature, obtained by applying a transverse field and assessing the amplitude of the $\mu^+$ precession generated. (c) Fast relaxation rates, $\lambda_F$, for all three samples and the highest oscillation frequency in the Lu sample, $\omega_1 = 2\pi f_1$, as functions of temperature. (d) Slow relaxation rate, or $1/T_1$, vs. temperature for all three samples.\label{C}}
\end{center}
\end{figure}

Specific heat, presented in Fig.~\ref{C}(a), shows peaks at 3.0 K, 5.2 K and 10.5 K for the In, Y and Lu samples, respectively, presumably indicating the onset of long range order (LRO) or spin freezing. First, it is quite clear that these materials are highly frustrated as the values of $\Theta_W$ we have determined are much higher than $T_m$, with the frustration likely arising from the triangular lattice geometry of the Ru dimers and the strong antiferromagnetic interactions between them. More specifically, we find frustration parameters, $f = \Theta_W/T_m$, of 13 (In), 21 (Y) and 11 (Lu). 

While our results are qualitatively consistent with previous work~\cite{Doi2002}, there is some variability in transition temperatures between our samples and those of Doi \emph{et al.}~\cite{Doi2002}. Whereas our $M$~$=$~Lu sample has a peak in the specific heat $C(T)$ at 10.5~K, their sample seems to have a 9.5~K ordering transition. The low-$T$ specific heat anomaly of our $M$~$=$~Y sample is also somewhat elevated when compared to Doi \emph{et al.}~\cite{Doi2002}. Meanwhile, our $M$~$=$~In sample has a peak in $C(T)$ that is broader and somewhat lower in temperature. Evidently there is some sample-dependence of the magnetic properties of these materials. 

\begin{figure}
\begin{center}
\includegraphics[width=3.3in,keepaspectratio=true]{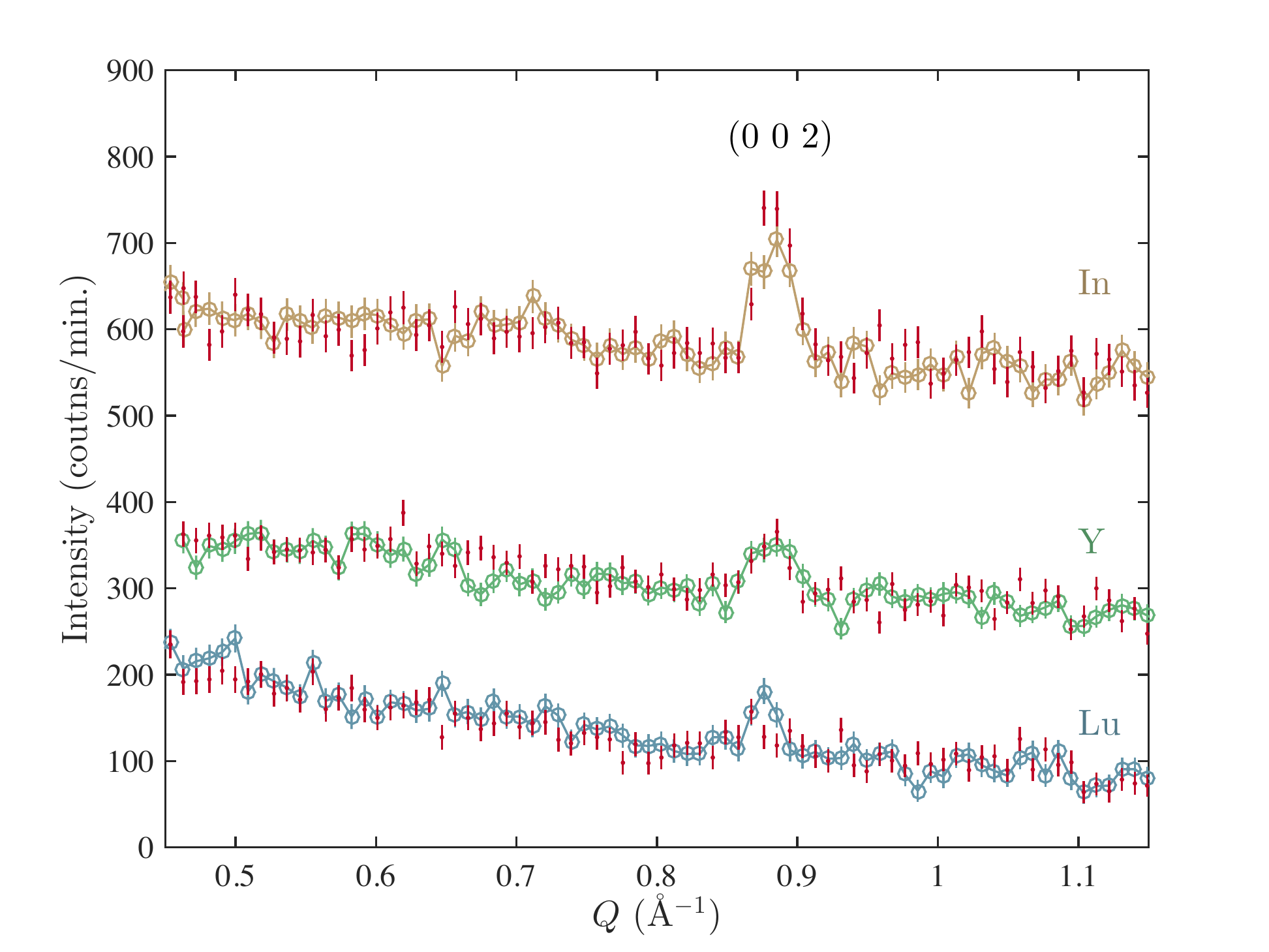}
\caption{Elastic neutron diffraction data from HB-1A for all three samples. The open circles are data taken at 1.5 K. Red points were taken above $T_m$, at 10 K (for Y and In) and 20 K for Lu. The Lu data has been shifted downwards by 200 counts / minute for ease of view. No evidence of magnetic order is observed, possibly because the ordered moment is extremely small or spread out over an entire dimer.\label{MagDiff}}
\end{center}
\end{figure}

Since there are possible indications of magnetic order or spin freezing in the $C(T)$ measurements, we performed elastic neutron scattering on the Ba$_3M$Ru$_2$O$_9$ ($M$~$=$~In, Y and Lu) samples using the HB-1A fixed incident energy triple axis spectrometer at HFIR of ORNL. The HB-1A experiment was designed to maximize the possibility of observing a magnetic signal, so this data is complementary to the HB-2A measurements described above where magnetic Bragg peaks were not observed. Specific advantages for the HB-1A experiment, as compared to the HB-2A measurements, are as follows: (1) The low-$T$ datasets were all measured at $T$~$=$~1.5~K to ensure that we were well below the $C(T)$ anomalies in each case. (2) The signal-to-noise at HB-1A relative to HB-2A is enhanced due to a double-bounce monochromator and the use of an analyzer for energy discrimination. Despite these improvements in the experimental set-up, the HB-1A measurements show no evidence of a magnetic signal below the $C(T)$ anomalies in each case, as illustrated in Fig.~\ref{MagDiff}. Although the HB-1A result for the Y sample appears to be inconsistent with previous work by Senn \emph{et al.}~\cite{Senn2013} using the WISH diffractometer at ISIS, it is important to note that the magnetic Bragg peaks observed in the WISH experiment were extremely weak. In fact, the ordered moment for the Y system reported in Ref.~\cite{Senn2013} is only 0.5(6)$\mu_B$ per Ru site, so there is a great deal of uncertainty in this value. The apparent discrepancy with the HB-1A data may simply arise due to a slightly different signal-to-noise ratio on WISH as compared to HB-1A, or there may be an extreme sensitivity of the Y magnetic ground state to some form of disorder.

Ref.~\cite{Senn2013} also reported the observation of significantly stronger magnetic Bragg peaks for Ba$_3$LaRu$_2$O$_9$. An ordered moment of $1.4(2)\mu_B$ per Ru site was determined from the subsequent magnetic refinement, which is consistent with a total spin of $S=3/2$ per dimer. Similar magnetic reflections were observed for both the La and Y samples, and this finding led the authors to conclude that these two materials host the same magnetic structure. Specifically, they find a (0 1/2 0) propagation vector which they attribute to a magnetic structure with ferromagnetic dimers. We note that their assumption of a ferromagnetic intradimer interaction is not consistent with our interpretation of the single dimer ground state for these materials, as discussed above. However, this discrepancy is resolved simply by replacing the single ion spins in their work with a single spin-1/2 moment distributed over each dimer in the case of the Y sample or a spin-3/2 moment in the case of La. The revised magnetic structure is then simply a collinear stripe phase, which has been predicted to arise for the quasi-2D triangular lattice when the NN and NNN in-plane exchange interactions are antiferromagnetic and comparable in magnitude \cite{Seabra2011}. These materials could therefore be considered to be the molecular magnet equivalents of isostructural compounds (for instance Ba$_3$CoSb$_2$O$_9$ \cite{Shirata2012,Zhou2012Co,Susuki2013,Koutroulakis2015,Quirion2015}) where the $M$-site is magnetic and forms a quasi-2d triangular lattice, albeit with a more important NNN interaction strength. 

\begin{figure}
\begin{center}
\includegraphics[width=3.5in,keepaspectratio=true]{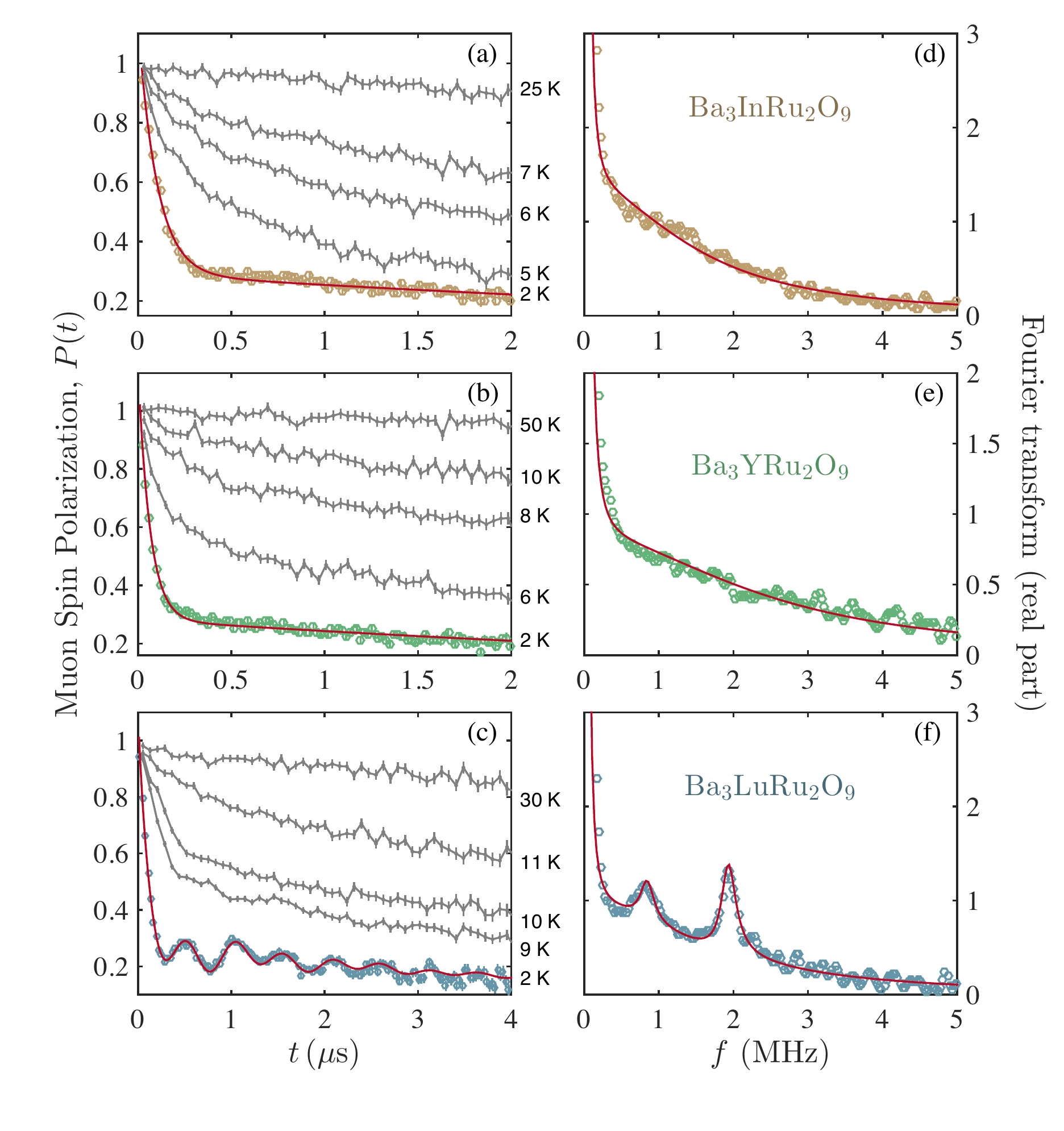}
\caption{Zero-field $\mu$SR asymmetry at various temperatures for (a) In, (b) Y and (c) Lu, with the corresponding fits superimposed on the 2~K data only. Fourier transforms of the 2~K data, with the fits superimposed, are shown in (d) In, (e) Y, and (f) Lu . The Lu sample data shows long-lived oscillations, whereas the data of the In and Y samples shows only fast exponential relaxation (along with a slowly relaxing 1/3 tail). The narrow zero-frequency peak in the Fourier transforms comes from the $T_1$ time of the 1/3 tail, and is a measure of spin fluctuations rather than static internal fields.\label{PandFT}}
\end{center}
\end{figure}

Since our neutron scattering measurements found no evidence for magnetic Bragg peaks in the In, Y, and Lu samples, possibly due to the small ordered moment sizes, it is natural to study these materials with $\mu$SR, which is one of the most sensitive probes of weak magnetism. As can be seen in Fig.~\ref{PandFT}, the ZF-$\mu$SR data of all three samples show indications of a magnetic phase transition with greatly increased relaxation at low $T$. The In and Y samples do not show any oscillations of the muon spin polarization thus the fast relaxation may arise from static disordered magnetism or slow spin fluctuations. As shown in Fig.~\ref{PandFT}(a) and (b), the muon spin polarization is well-described by the following two-component relaxation function 
\begin{equation}\label{DblExp}
P(t) = (1-x)e^{-\lambda_f t} + x e^{-t/T_1} 
\end{equation}
where $x$ is close to 1/3 at low $T$ and $1/T_1$ is the spin-lattice relaxation rate~\cite{Colman2011}. Assuming that we are in the quasi-static limit, $\lambda_f$ results largely from inhomogeneities (disorder) in the static internal fields at the muon site(s), whereas $1/T_1$ is caused by residual spin fluctuations.

As shown in Fig.~\ref{PandFT}(c), clear oscillations of the polarization are observed in the low $T$ regime for the Lu sample. The Fourier transform of this data, illustrated in Fig.~\ref{PandFT}(f), shows two distinct frequencies corresponding to rather small internal fields of 6.1(2) and 14.1(1) mT. The two frequencies are indicative of two distinct muon stopping sites, which can likely be associated with the two crystallographically-inequivalent oxygen atoms in the crystal structure. It is also possible that one crystallographic muon stopping site could give rise to two distinct frequencies as a result of a complex magnetic structure. However, the magnetic structure reported by Senn \emph{et al.}~\cite{Senn2013} should only lead to one frequency per crystallographic site, so in that particular case our spectrum would arise from two crystallographically distinct muon stopping sites.

The Fourier transform also shows that these two peaks are superimposed on a broad feature, which is consistent with the fast exponential relaxation observed in the time domain. Hence, the Lu data can be fit with the following equation :
\begin{equation}\label{PLu}
P(t) = (1-x) \sum_{n=0}^2 a_n \cos(2\pi f_n t) e^{-\lambda_n t} + x e^{-t/T_1} 
\end{equation}
where $f_0$~$=$~0 and $\lambda_0 = \lambda_f = 10.9(7)$ $\mu$s$^{-1}$ is the fast relaxing exponential component. Despite the fact that the oscillations in the muon spin polarization are very well-resolved, our fits reveal that they come from a relatively small portion of the sample, 15\%, with the remainder of the sample behaving more similarly to the In and Y analogs.  The fitting parameters obtained in ZF at the lowest temperatures are presented in Table~\ref{ParamTable}.

TF-$\mu$SR measurements (in a field of $\sim$50 G) were used to rapidly map out the transitions. The data was fit with the following equation:
\begin{equation} \label{PTF}
P(t) = f_\mathrm{PM}\cos(\gamma B_\mathrm{TF} t + \phi) e^{-\lambda t} + (1-f_\mathrm{PM})
\end{equation}
where $f_\mathrm{PM}$, shown as a function of temperature in Fig.~\ref{C}(b), is the fraction of the sample that remains paramagnetic (and therefore has oscillations of the muon spin polarization induced by the applied magnetic field). The other fraction of the sample hosts either static magnetism or strong spin dynamics that dwarf the small applied transverse field. It is interesting to compare the temperature evolution of the paramagnetic fraction to the specific heat, the maximum of which can be taken as the transition temperature, $T_m$. For the Lu sample, $f_\mathrm{PM}$ begins to drop below 100\%~precisely at $T_m$. On the other hand, the paramagnetic volume fraction deviates from 100\%~well above $T_m$ for the In and Y samples, which suggests that there is a broad temperature regime of short-range magnetic order. 

\begin{table}[h]
\caption{Various experimental parameters for the three samples studied.  The transition temperature, $T_M$, is obtained from the maximum in specific heat. The first energy gap to the $S=3/2$ excited state, $\Delta_1$, is obtained from inelastic neutron scattering. The magnetic susceptibility, allows for a determination of the Weiss constant, $\Theta_W$, the effective moment, $\mu_\mathrm{eff}$, and the second energy gap to the $S=5/2$ excited state, $\Delta_2$. From $\mu$SR, the muon oscillation frequencies, $f_1$ and $f_2$, the corresponding line widths, $\lambda_1$ and $\lambda_2$ and the fast relaxation rate, $\lambda_f$, are presented. \label{ParamTable} }
\begin{center}
\begin{tabular}{lllll}
\hline
\hline
Technique & Parameter & In & Y & Lu \\
\hline

$C$ & $T_M$ (K) &   				3.0(3) & 5.2(1) & 10.5(2) \\
\vspace{-0.05in}\\
\hline
INS & $\Delta_1$ (meV) & 	34.0(1.0)	& 31.5(1.5) & 34.0(1.0) \\
\vspace{-0.05in}\\
\hline
$\chi$ & $\Delta_2$ (meV) &  81(1) & 72(1) & 80(1) \\
&$\Theta_W$ (K) &			43(3) & 110(10) & 113(2) \\
&$\mu_\mathrm{eff}/\mu_B$ &	1.40(3) & 1.65(3) & 1.53(3) \\
\vspace{-0.05in}\\
\hline
$\mu$SR & $\lambda_f$ ($\mu$s$^{-1}$) &  9.9(3) & 15.7(6) & 10.9(7) \\
\vspace{-0.05in}\\
& $f_1$ (MHz) &   		         -- & -- & 0.83(3) \\
& $\lambda_1$ ($\mu$s$^{-1}$) & -- & -- & 0.7(4) \\
\vspace{-0.05in}\\
&$f_2$ (MHz) &				 -- & -- & 1.91(2) \\				
&$\lambda_2$ ($\mu$s$^{-1}$) & -- & -- & 0.74(16) \\
\hline
\hline
\end{tabular}
\end{center}
\label{default}
\end{table}%

Performing ZF-$\mu$SR measurements as a function of $T$ has allowed us to extract the temperature dependence of $1/T_1$, as well as the fast relaxation rate, $\lambda_f$. In the case of the Lu system, we can also track one of the precession frequencies, $f_1$, as a function of temperature, whereas the lower frequency, $f_2$, is only quantifiable at the lowest temperatures. These results are shown in panels (c) and (d) of Fig. \ref{C}.  $f_1(T)$ develops rather sharply at the Lu transition temperature and the $T$-dependence resembles a standard order-parameter plot. On the other hand, $\lambda_f$ evolves very gradually for all three samples with no sharp change at $T_m$. $1/T_1$ shows a peak near 4 K in the data for the Y and In samples, which is typical of critical spin dynamics. In the case of the Lu sample, there is a much weaker and broader feature in $1/T_1$.

\begin{figure}
\begin{center}
\includegraphics[width=3.3in,keepaspectratio=true]{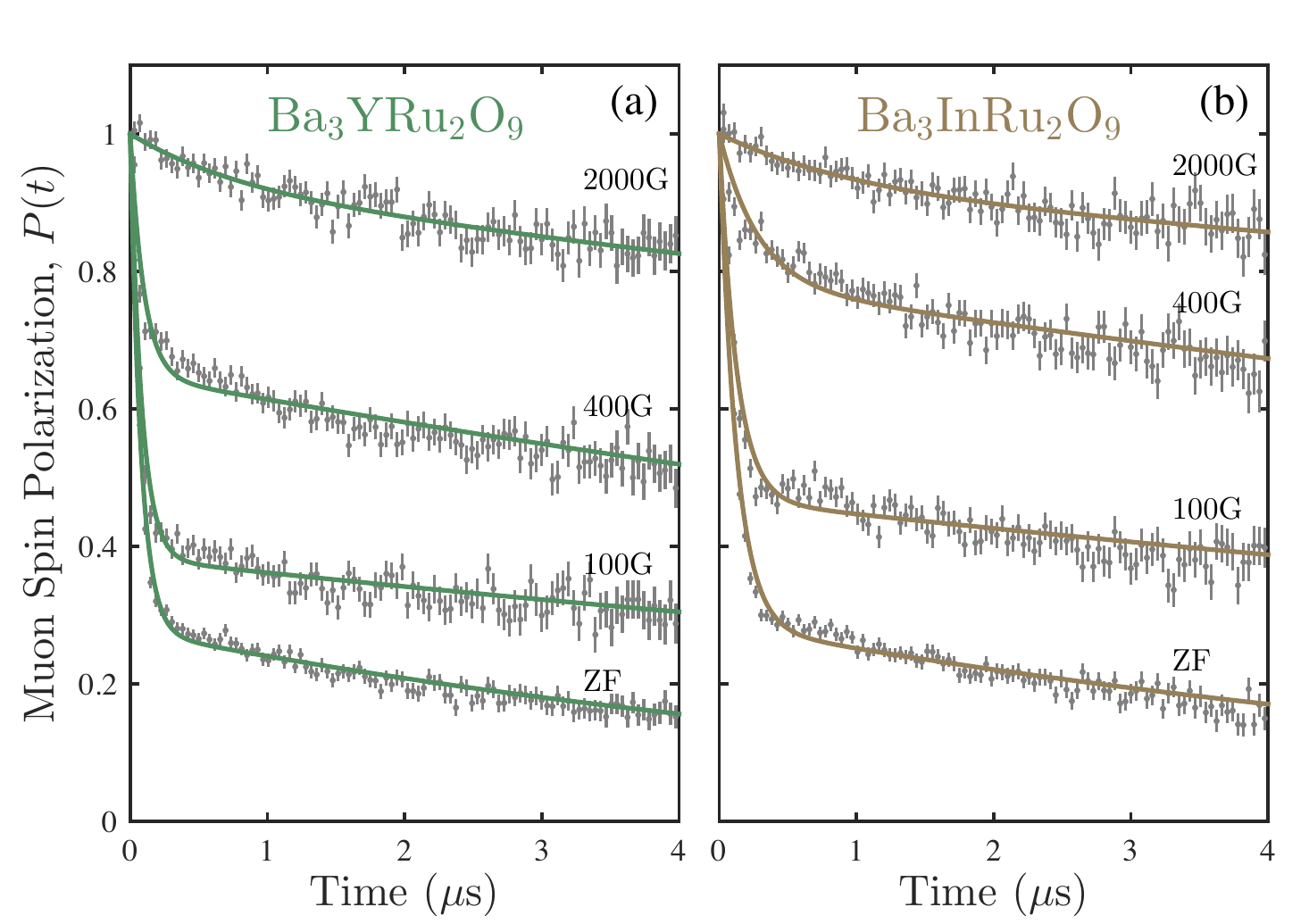}
\caption{Longitudinal field $\mu$SR scans for the Y and In samples.\label{LF}}
\end{center}
\end{figure}

The two-component exponential relaxation observed in the Y and In samples can be interpreted in two ways. First, in a quasi-static picture, the slow relaxation arises from a so-called 1/3 tail with a weak $1/T_1$ relaxation rate coming from residual spin fluctuations, and the fast relaxation is the 2/3 component coming from random internal fields. Alternatively, one could suspect a dynamic, but inhomogeneous, material with two different $T_1$ times. Longitudinal field scans at the lowest temperature, shown in Fig.~\ref{LF}, confirm that the fast relaxation is a result of static inhomogeneities as it is decoupled fairly quickly. More precisely, in the Y sample, the fast relaxation is $\lambda_f = 15.7(6)$ $\mu$s$^{-1}$, implying an internal field distribution of width $\Delta B \simeq \lambda_f/\gamma_\mu = 184(7)$ G. Thus, the application of a longitudinal field equal to $B_\mathrm{LF} = 10\Delta B$, should entirely decouple the muon spins from the internal field and eliminate the fast-relaxing 2/3 component of $P(t)$~\cite{YaouancBook}. As shown in Fig.~\ref{LF}(a), this appears to be the case for the LF~$=$~2000 G spectrum. Furthermore, as seen in Fig.~\ref{LF}(b), the ZF fast relaxation for the In sample ($\lambda_f = 9.9(3)$ $\mu$s$^{-1}$) is somewhat more easily decoupled via application of a longitudinal field, as expected. It is thus clear that these materials host static magnetic ground states from the perspective of $\mu$SR.

\begin{figure}
\begin{center}
\includegraphics[width=3.3in,keepaspectratio=true]{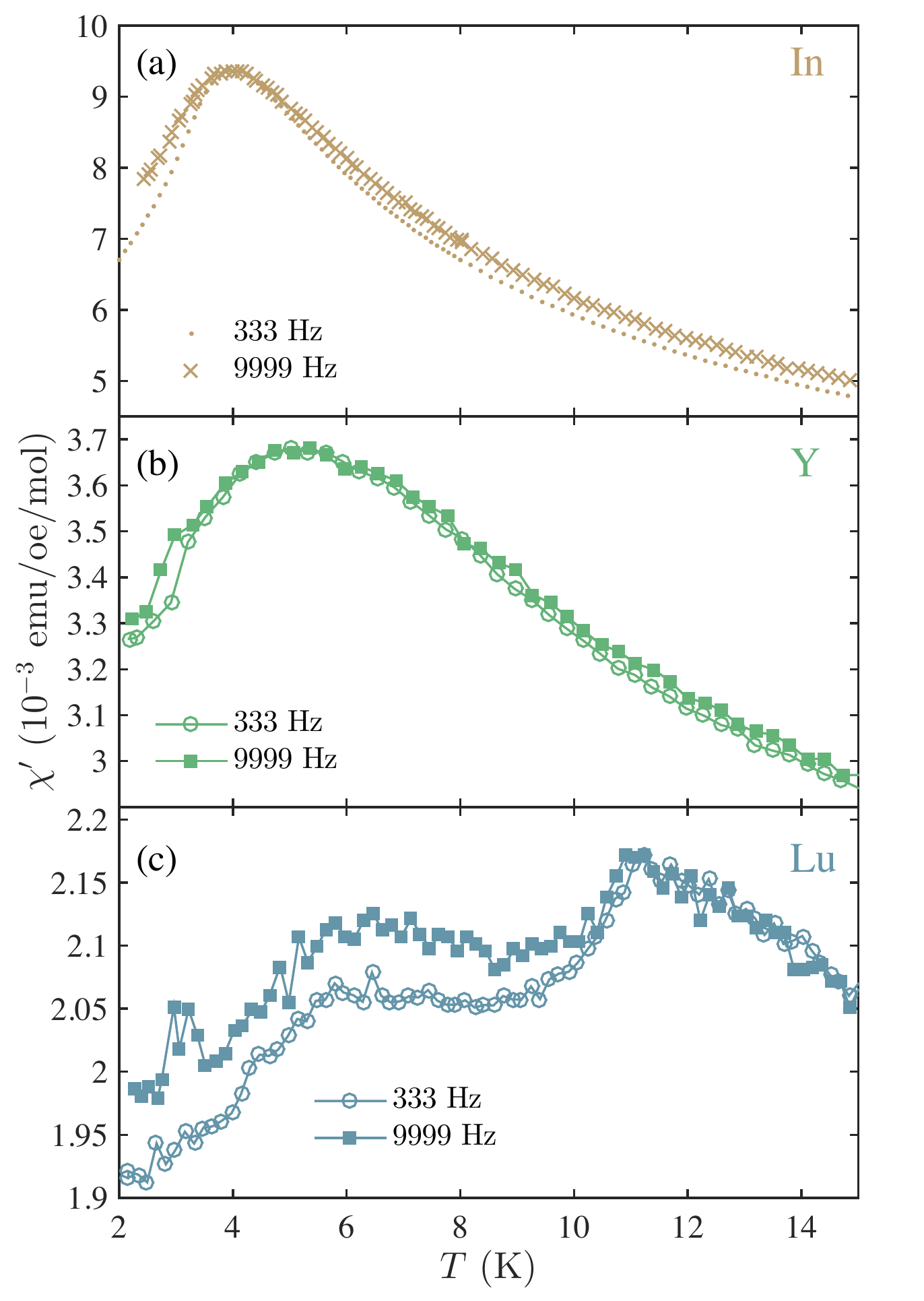}
\caption{AC susceptibility ($\chi'$) of all three samples at two different frequencies, 333 Hz and 9999 Hz. The high frequency data has been normalized so that the peak susceptibilities are equal, since the frequency response of the \emph{PPMS} system is not perfectly flat. Otherwise, the temperature dependence of the susceptibilities are very similar. In particular, the peak positions are independent of frequency. \label{acChi}}
\end{center}
\end{figure}

It is tempting to attribute the lack of oscillations in the ZF muon spin polarization of the In and Y samples to spin glass physics, especially since a zero-field-cooled/field-cooled divergence has been previously observed at $T_m$ in the DC susceptibility of the former system \cite{Shylk2007}. Furthermore, many geometrically-frustrated magnetic materials show a strong sensitivity to tiny amounts of quenched crystalline disorder which can lead to a spin glass transition~\cite{Gingras1997,Schiffer1995,QuilliamGarnets,Bisson2008}. However, we have also measured the AC susceptibility of these materials at several different frequencies (ranging from 333 Hz to 9999 Hz) and found no evidence of spin glassiness. More specifically, as shown in Fig.~\ref{acChi}, the position, $T_\mathrm{max}$, of the real part of the ac susceptibility, $\chi'(T)$, is independent of frequency, in the frequency range studied. A conventional spin glass will show a maximum in $\chi'$ at the freezing temperature, $T_f$, which then depends strongly on the frequency of measurement, with an extrapolation to zero-frequency allowing for a determination of the true glass temperature, $T_g$~\cite{Paulsen1987}. Whereas the In and Y samples have a single peak in $\chi'$, the Lu sample has a somewhat more complicated susceptibility, with a relatively sharp peak at $\sim 11$ K, corresponding to the peak in specific heat and the onset of oscillations in $\mu$SR and a lower-temperature peak, similar to that of the Y sample, which likely corresponds to the gradual onset of fast relaxation ($\lambda_f$) in the $\mu$SR spectra. In other words, the broad, lower temperature peak, is associated with the disordered portion of the sample. Nonetheless, this peak does not seem to show an appreciable dependence on frequency, but simply a very slight increase in magnitude at 9999 Hz. These two features end up forming a rather broad critical temperature regime which is consistent with the broad $1/T_1$ feature observed in our $\mu$SR experiments on the Lu sample. Given our AC susceptibility results and the fact that magnetic Bragg peaks were observed in neutron diffraction measurements on a different Y sample \cite{Senn2013}, it appears that these materials  are not conventional spin glasses and likely have long-range ordered ground states.

There are several possible origins for strongly-damped oscillations in the $\mu$SR data. We will concentrate on static origins only, since the well-defined 1/3 tail in our data indicates that the spins are mostly static, or fluctuating so slowly that they are essentially static from the point of view of $\mu$SR. The two possible static origins of the strong damping are (1) a large number of inequivalent muon stopping sites and (2) a modulation of the internal fields by disorder. The first scenario is highly unlikely given the two well-defined oscillations in the Lu data, which imply that there are two preferred crystallographic sites for the muons. On the other hand, the second scenario appears to be compatible with our $\mu$SR and neutron scattering results. An antiferromagnetic, symmetry-breaking long-range order can coexist with a large random modulation of the moments. This large amount of disorder can lead to the loss of oscillations in the ZF muon spin polarization and a reduced magnetic signal in neutron scattering that is not observed in our measurements. Given the discrepancy between our results and earlier neutron diffraction work on the Y system~\cite{Senn2013}, it is logical to suspect the influence of sample-dependent disorder  on the magnetic ground state.

It is also valuable to consider the implications of the observed $\mu$SR signals for the molecular magnet model proposed above, notably through the size of the measured internal fields. Dipolar coupling to point-like dipoles of $0.5\mu_B$ per site ($S=1/2$ per dimer) should give rise to an oscillation frequency of $\sim 7$ MHz for a $\mu^+$ stopping $\sim1\,\mathrm{\AA}$ away from the O$_1$ site. Hence, the fact that we observe $f_1 = 1.91(2)$ MHz in the Lu sample implies a magnetic moment of only $0.14\mu_B$ ($0.28\mu_B$ per dimer). Evidently a model of point-like dipoles on the Ru sites is highly simplistic. Even so, our results indicate that the spins are probably very much extended over an entire Ru$_2$O$_9$ ``molecule'' which is consistent with the orbital hybridization picture discussed above. Indeed, the slow oscillations seen here resemble those observed in molecular magnets where each spin is distributed over an entire molecular unit~\cite{Le1993, Blundell2004}. The ordered magnetic moments for the Y and In samples appear to be similarly weak, which is likely why no magnetic signal was detected in our elastic neutron diffraction measurements. 

Finally, we can speculate as to why the Y and In samples show such a high level of disordered static magnetism. As can be seen in Fig.~\ref{Orbitals}(a) and (b), the $e_g^\pi$ and $e_g^{\pi\ast}$ orbitals remain degenerate in the Ba$_3M$Ru$_2$O$_9$ structure. For the Lu, Y and In samples, only one electron occupies the anti-bonding $e_g^{\pi\ast}$ orbitals and therefore they are Jahn-Teller (JT) active. Importantly, this degeneracy is not lifted by the spin-orbit coupling~\cite{Kugel2015}. This may leave these materials vulnerable to local structural distortions that relieve the degeneracy, but lead to disorder in the interdimer exchange or the crystalline electric field, both of which can modulate the size of the ordered moments. An important parallel can be found in the sister compound Ba$_3$CuSb$_2$O$_9$, which is also based on $S=1/2$ moments and Jahn-Teller active~\cite{Nakatsuji2012}. In Ba$_3$CuSb$_2$O$_9$, two distinct behaviors are observed, depending on the precise stoichiometry of the samples~\cite{Nakatsuji2012,Katayama2015}. In some off-stoichiometric samples, the orbital degeneracy is relieved by an orthorhombic distortion (a collective JT transition) near 200 K. Ultimately, these orthorhombic samples order magnetically at low temperatures. More stoichiometric samples manage to preserve their room-temperature hexagonal symmetry down to much lower temperatures either through a dynamic JT effect~\cite{Ishiguro2013} or else local distortions that nonetheless preserve the global symmetry of the structure and give rise to a random-singlet magnetic ground state~\cite{Quilliam2012bcso}. The most recent experimental results, thermal conductivity measurements, on nearly stoichiometric single-crystal Ba$_3$CuSb$_2$O$_9$ point toward the local-distortion picture~\cite{Sugii2016} which is consistent with the random singlet magnetic ground state and excitation gap~\cite{Quilliam2012bcso}. Since a hexagonal to orthorhombic collective JT transition can be ruled out by the neutron diffraction results on the materials studied here, similar random distortions might then apply, and they may be extremely important for understanding the collective magnetic ground states and possible sample dependence of the magnetic properties. Future work should search for these local distortions, possibly via x-ray absorption fine structure measurements. 

\section{VI. Conclusions}
We have used a wide array of experimental techniques to characterize both the single dimer and collective magnetic properties of the mixed valence Ru dimer systems Ba$_3M$Ru$_2$O$_9$ ($M$~$=$~In, Y and Lu). Our combined neutron powder diffraction, DC magnetic susceptibility, and inelastic neutron scattering results indicate that the Ru dimers are best described as molecular units with one spin-1/2 moment distributed equally over the two Ru sites. Two dispersive magnetic excitations are observed in the inelastic neutron scattering spectrum of each system. We attribute the lower energy mode to electron transitions between antibonding orbitals, while the upper mode is argued to arise from electron transitions between bonding and antibonding orbitals.

The dimers form a quasi-2D triangular lattice, which is strongly frustrated due to significant antiferromagnetic interdimer exchange. Our heat capacity and muon spin relaxation results reveal that the molecular moments develop a static magnetic ground state in each case, with clear evidence of long-range magnetic order for the Lu sample. The size of the static internal fields observed in $\mu$SR at low temperatures are consistent with $S=1/2$ moments distributed over an entire Ru$_2$O$_9$ dimer, similar to molecular magnets.  Although the static magnetism is much more disordered for the Y and In samples, they do not appear to be conventional spin glasses, for example. Overall, the current work demonstrates that the 6H-perovskites Ba$_3MA_2$O$_9$ are excellent model systems for detailed investigations of frustrated quantum magnetism arising from spin-1/2 molecular building blocks on a triangular lattice.  Given the strong theoretical interest in $S=1/2$ triangular-lattice antiferromagnets and the rarity of representative materials, these systems should be attractive for future studies of the magnetic ground state and magnetization process, albeit with the added complexity of orbital degrees of freedom. Finally, we note that our results can likely be directly applied to understanding the magnetic properties of the related Ir-dimer system, Ba$_3$InIr$_2$O$_9$, which also seem to be consistent with spin-1/2 dimers at low temperature, and moreover appear to indicate a gapless quantum spin liquid ground state~\cite{Dey2017}.

\begin{acknowledgments}

We are grateful to the staff of the Centre for Molecular and Materials Science at TRIUMF for extensive technical support, in particular G. Morris, B. Hitti,  D. Arseneau, and I. MacKenzie. We acknowledge useful conversations with S. Johnston and G.E. Granroth and are particularly grateful to S. Streltsov for helping us to understand the intradimer physics of these systems. A portion of this research used resources at the High Flux Isotope Reactor and Spallation Neutron Source, which are DOE Office of Science User Facilities operated by Oak Ridge National Laboratory. J. Q. acknowledges research funding obtained from NSERC and the FRQNT. R.S. and H. Z. acknlowedge the support of NSF-DMR-1350002. C. Q. acknowledges the CEM, and NSF MRSEC, under Grant No. DMR-1420451. 
\end{acknowledgments}


\end{document}